# CLIC – A TWO-BEAM MULTI-TeV e$^{\pm}$ LINEAR COLLIDER

J.P. Delahaye and I. Wilson for the CLIC study team:
R. Assmann, F. Becker, R. Bossart, H. Braun, H. Burkhardt, G. Carron, W. Coosemans, R. Corsini, E. D'Amico, S. Doebert, S. Fartoukh, A. Ferrari, G. Geschonke, J.C. Godot, L. Groening, G. Guignard, S. Hutchins, B. Jeanneret, E. Jensen, J. Jowett, T. Kamitani, A. Millich, P. Pearce, F. Perriollat, R. Pittin, J.P. Potier, A. Riche, L. Rinolfi, T. Risselada, P. Royer, F. Ruggiero, D. Schulte, G. Suberlucq, I. Syratchev, L. Thorndahl, H. Trautner, A. Verdier, W. Wuensch, F. Zhou, F. Zimmermann, CERN, Geneva, Switzerland, O. Napoly , SACLAY, France.

## ABSTRACT

The CLIC study of a high-energy (0.5 - 5 TeV), high-luminosity ($10^{34}$ - $10^{35}$ cm$^{-2}$ sec$^{-1}$) e$^{\pm}$ linear collider is presented. Beam acceleration using high frequency (30 GHz) normal-conducting structures operating at high accelerating fields (150 MV/m) significantly reduces the length and, in consequence, the cost of the linac. Using parameters derived from general scaling laws for linear colliders, the beam stability is shown to be similar to lower frequency designs in spite of the strong wake-field dependency on frequency. A new cost-effective and efficient drive beam generation scheme for RF power production by the so-called "Two-Beam Acceleration" method is described. It uses a thermionic gun and a fully-loaded normal-conducting linac operating at low frequency (937 MHz) to generate and accelerate the drive beam bunches, and RF multiplication by funnelling in compressor rings to produce the desired bunch structure. Recent 30 GHz hardware developments and CLIC Test Facility (CTF) results are described.

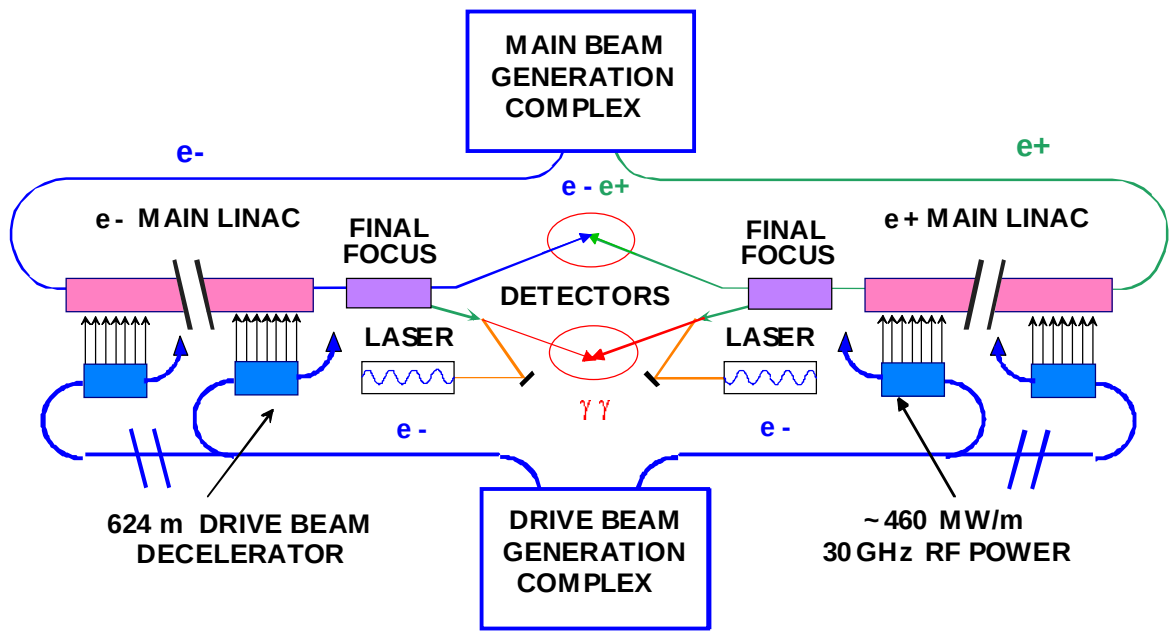

**Figure 1: Overall layout of the CLIC complex.**

## 1 INTRODUCTION

The Compact Linear Collider (CLIC) covers a centre-of-mass energy range for e$^{\pm}$ collisions of 0.5 - 5 TeV [1] with a maximum energy well above those presently being proposed for any other linear collider [2]. It has been optimised for a 3 TeV e$^{\pm}$ colliding beam energy to meet post-LHC physics requirements [3] but can be built in stages without major modifications. An overall layout of the complex is shown in Fig.1. In order to limit the overall length, high accelerating fields are mandatory and these can only be obtained with conventional structures, by operating at a high frequency. The RF power to feed the accelerating structures is extracted by transfer structures from high-intensity/low-energy drive beams running parallel to the main beam (Fig. 2). A single tunnel, housing both linacs and the various beam transfer lines without any modulators or klystrons, results in a very simple, cost effective and easily extendable configuration.

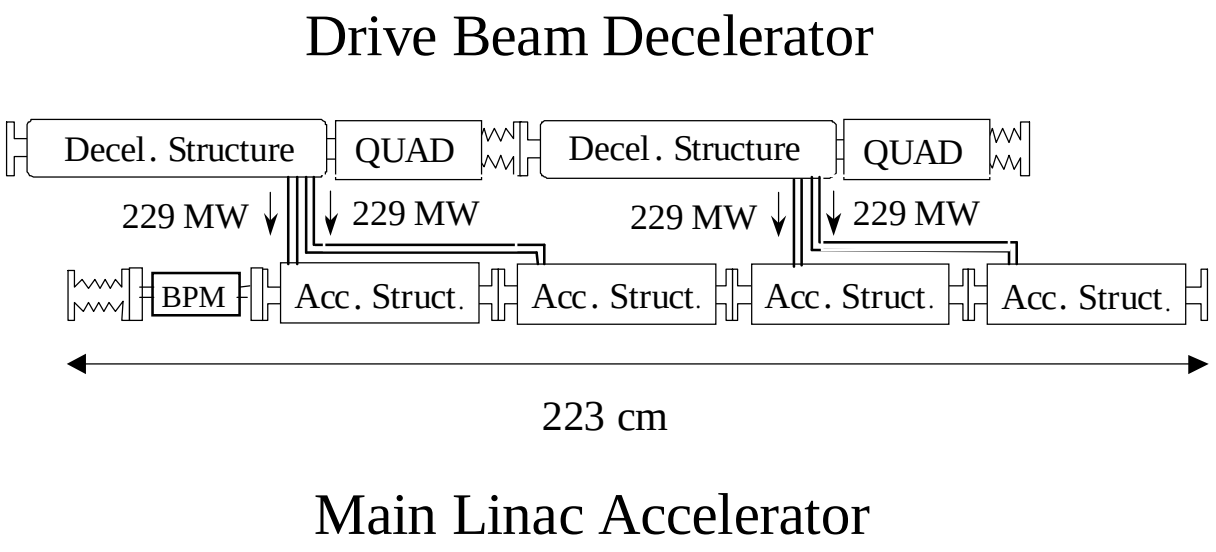

**Figure 2: One main-beam and drive-beam module.**

## 2 MAIN PARAMETERS

The main-beam and linac parameters are listed in Table 1 for two colliding beam energies. The luminosity $L$ normalised to the RF power, $P_{RF}$, depends on a small number of parameters in both low ($\Upsilon$<<1) and high beamstrahlung ($\Upsilon$>>1) regimes:

$$\frac{L_{\Upsilon<<1}}{P_{RF}} \propto \frac{\delta_B^{1/2}}{U_f}\frac{\eta_b^{RF}}{\varepsilon_{ny}^{*1/2}} \text{ and } \frac{L_{\Upsilon>>1}}{P_{RF}} \propto \frac{\delta_B^{3/2}}{U_f^{1/2}\beta_y^{*1/2}}\frac{\eta_b^{RF}}{\sigma_z^{1/2}\varepsilon_{ny}^{*1/2}} \quad (1)$$

where $\delta_B$, is the mean energy loss, $\eta^{RF}_b$ the RF-to-beam efficiency and $U_f$, $\sigma_z$, $\beta_y$, $\varepsilon^*_{ny}$ the beam energy, bunch length, vertical beta function and normalised vertical beam emittance at the I.P. respectively [4]. The parameters have been derived from general scaling laws [4] covering more than a decade in frequency. These scaling laws, which agree with optimised linear collider designs, show that the beam blow-up during acceleration can be made independent of frequency for equivalent beam trajectory correction techniques. As a consequence, and in spite of the strong dependence of wakefields on frequency, CLIC whilst operating at a high frequency but with a low charge per bunch $N$, a short bunch length $\sigma_z$, strong focussing optics and a high accelerating gradient $G$, preserves the vertical emittance as well as low frequency linacs. The RF-to-beam transfer efficiency is optimised by using a large number of bunches and by choosing an optimum accelerating section length. In spite of the reduced charge per bunch and the high gradient, excellent RF-to-beam efficiency is obtained because the time between bunches is shorter and the shunt impedance of the accelerating structures is higher.

**Table 1: Main beam and linac parameters**

| Beam parameters at IP | 1 TeV | 3 TeV |
|---|---|---|
| Luminosity ($10^{34}cm^{-1}s^{-1}$) | 2.7 | 10.0 |
| Mean energy loss (%) | 11.2 | 31 |
| Photons /electrons | 1.1 | 2.3 |
| Coherent pairs per crossing | $3\times10^6$ | $6.8\times10^8$ |
| Repetition rate (Hz) | 150 | 100 |
| $e^\pm$ / bunch | $4\times10^9$ | $4\times10^9$ |
| Bunches / pulse | 154 | 154 |
| Bunch spacing (cm) | 20 | 20 |
| $\varepsilon_n$ ($10^{-8}$ rad.m) H/V | 130/2 | 68/2 |
| Beam size (nm) H/V | 115/1.75 | 43/1 |
| Bunch length (μm) | 30 | 30 |
| Accel. gradient (MV/m) | 150 | 150 |
| Two-linac length (km) | 10 | 27.5 |
| Accelerating structures | 14140 | 42940 |
| Power / section (MW) | 229 | 229 |
| Number of 50 MW klystrons | 364 | 364 |
| Klystron pulse length (μs) | 33.3 | 92 |
| RF-to-beam efficiency (%) | 24.4 | 24.4 |
| AC to beam efficiency (%) | 9.8 | 9.8 |
| AC power (MW) | 150 | 300 |

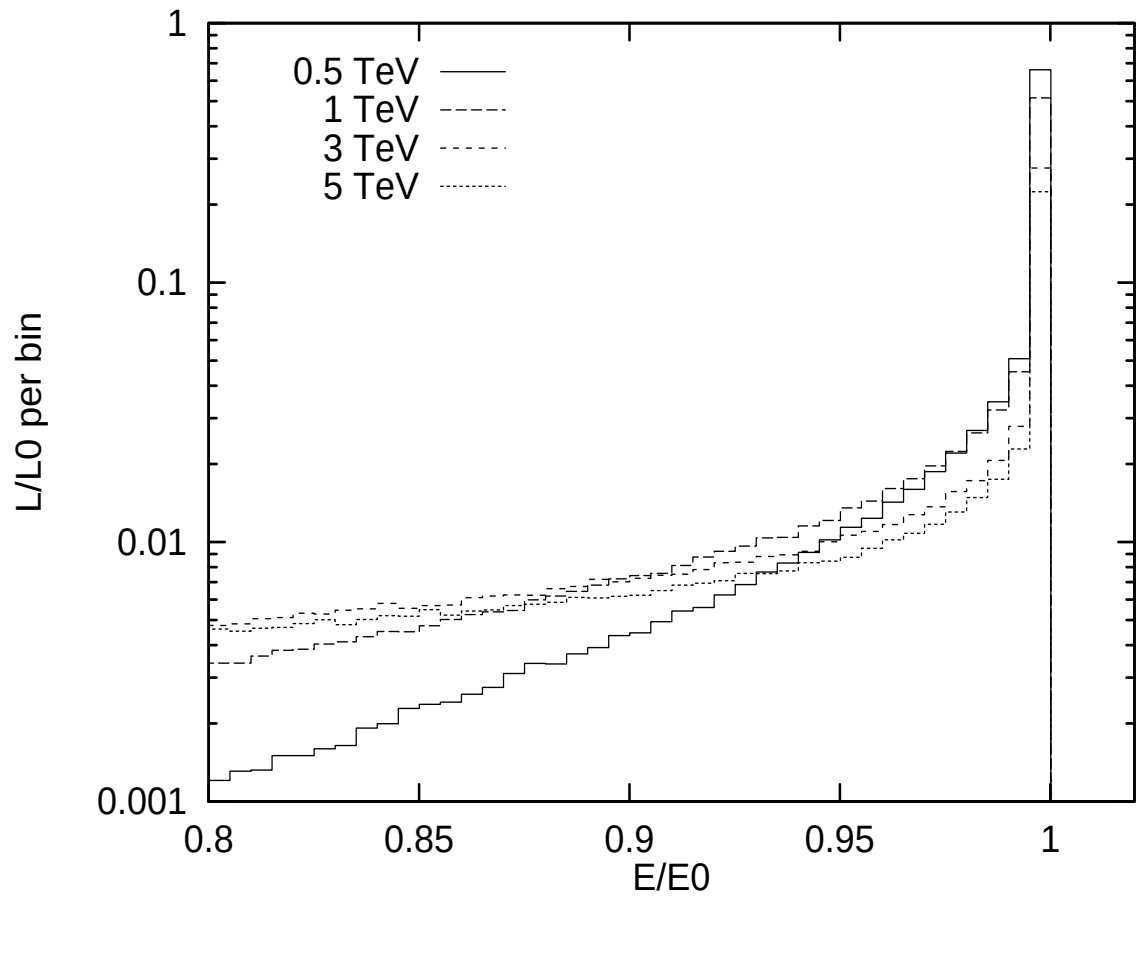

**Figure 3: Luminosity distribution with energy. (L0: total integrated luminosity, E0: max. energy)**

Up to 1 TeV, where the beamstrahlung parameter $Y$<1, the beam parameters are chosen to have a small $\delta_B$. To limit the power consumption above 1 TeV, $\varepsilon_{ny}$ is reduced and $Y$ allowed to be >>1. In this regime (see Eq.1), high frequency linacs are very favourable because $\sigma_z$ is small. As a consequence, even with $Y$ >>1, neither the $L$ spectrum (Fig. 3), nor the number of emitted gammas which increase the background in the detector, significantly deteriorate with energy [1] (see Table 1). The number of $e^\pm$ pairs generated per crossing however increases significantly with energy.

# 3 MAIN LINAC

The effects of the strong 30 GHz wakefields ($W_T$) can be kept moderate by choosing $N$ to be small ($4\times10^9$ at all energies) and $\sigma_z$ at the lower limit that is permitted by the momentum acceptance of the final focus. With a high gradient $G$ and strong focusing, the single-bunch blow-up $\Delta\varepsilon_{ny}$ can be kept below ≈100 % at all energies (Fig. 4) [5].

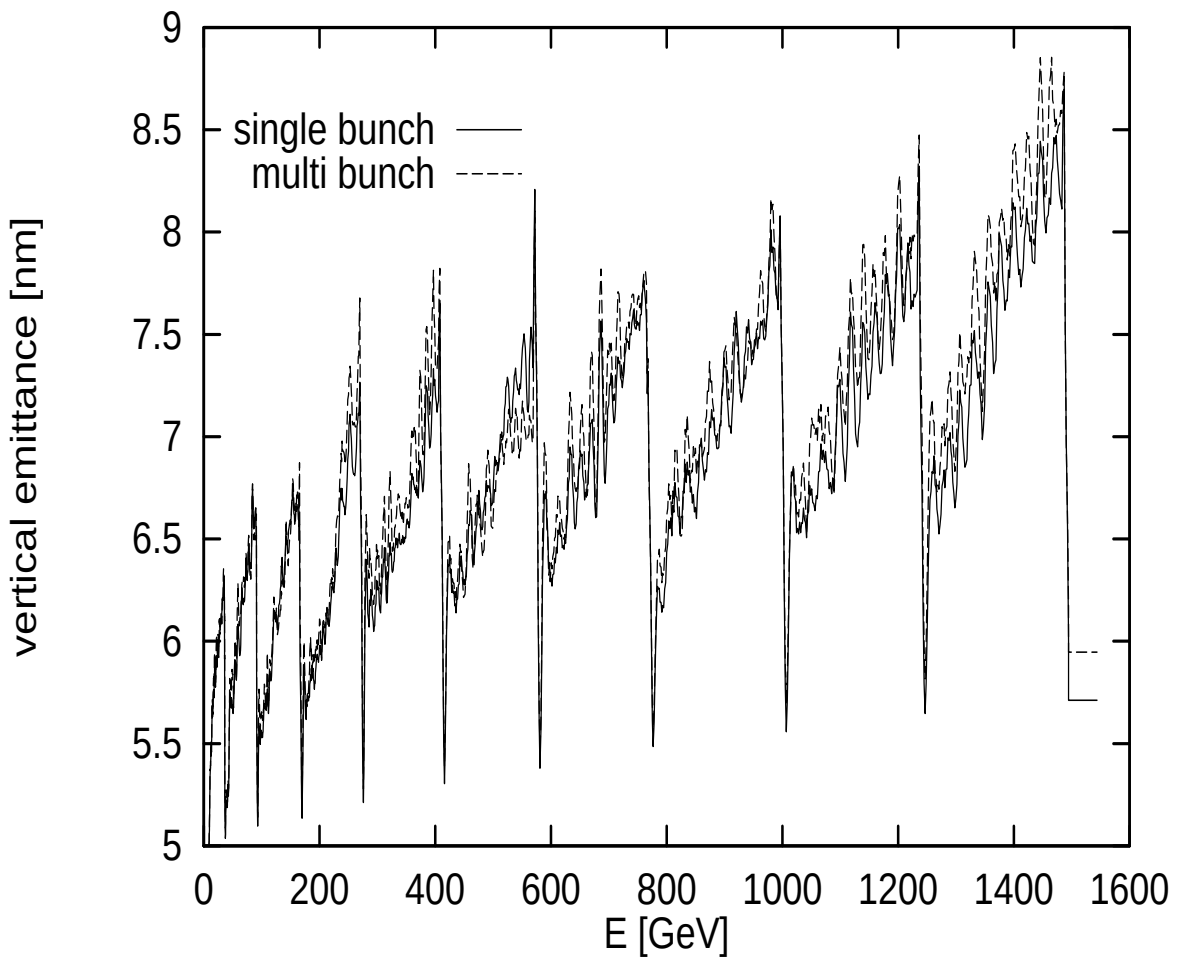

**Figure 4: Emittance variation along the main linac.**

To obtain the values of $L$ given in Table 1 a very small injected $\varepsilon_{ny}$ of $5\times10^{-9}$ rad.m is assumed. Limiting the overall $\Delta\varepsilon_{ny}$ relies in part on the use of bumps which are created locally at 5-10 positions along the linac by mis-aligning a few upstream cavities. The effects of these bumps are used to minimise the local $\varepsilon_{ny}$ (Fig.4). Without these bumps, dispersive effects are ≈10 times weaker than $W_T$ effects. The average lattice β-function starts from ≈ 4-5 m and is scaled approx. as $(\text{energy})^{0.5}$. The FODO lattice is made up of sectors with equi-spaced quadrupoles of equal length and normalised strength, with matching insertions between sectors. The RF cavities and quadrupoles are pre-aligned to 10 and 50 μm respectively using a stretched-wire positioning system. The misalignments of the beam position monitors (BPMs) are measured as follows [6]. A section of 12 quadrupoles is switched off, and with the beam centred in the two end BPMs of this section, the relative mis-alignment of the other monitors are measured with an accuracy of 0.1μm. The beam trajectory and ground motion effects are corrected by a 1-to-1 correction. BNS damping is achieved by running off the RF-crest by $6^o$ to $10^o$. Multiple bunches are required to obtain high luminosities. The multi-bunch emittance blow-up $\Delta\varepsilon_{ny}$ is ≈ 20%. To make the 154-bunch train stable requires a strong reduction of the transverse wakefields induced by the beam in the accelerating structures. A new Tapered Damped Structure (TDS) [7] has been designed. Each of the 150 cells is damped by its own set of four radial waveguides (Fig. 5) giving a $Q$ of 16 for the lowest dipole mode. A simple linear tapering of the iris dimension provides a de-tuning frequency spread of 2 GHz (5.4%). The waveguides are terminated with short silicon carbide loads [8].

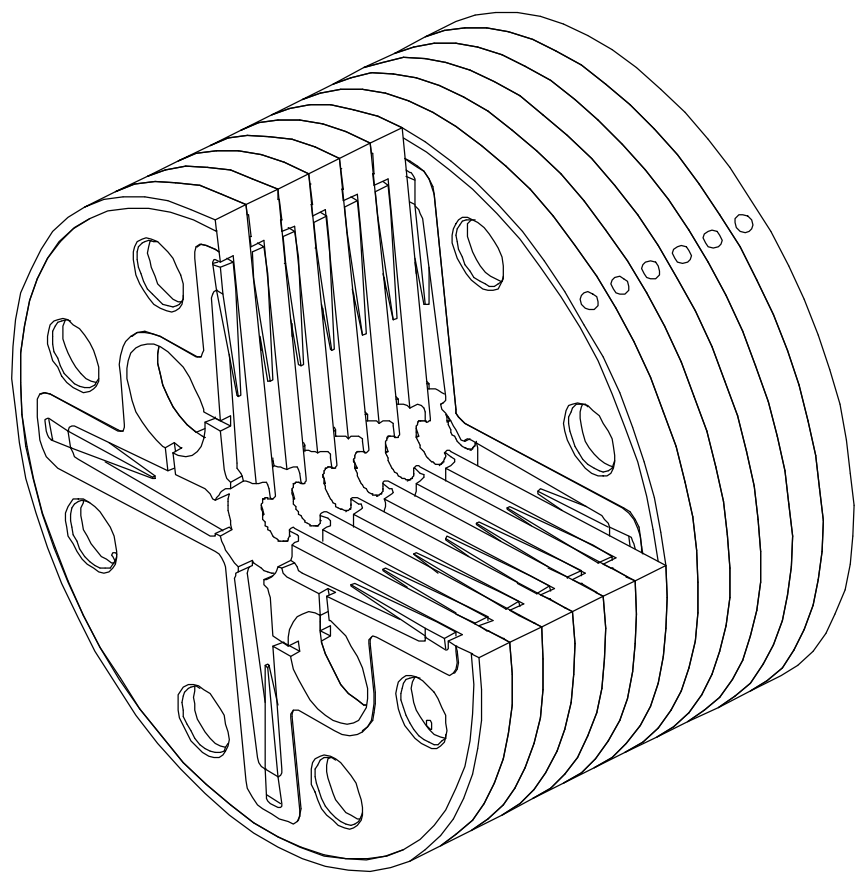

**Figure 5: A cut-away view of the CLIC TDS**

Calculations of the transverse wakefields in this structure indicate a short-range level of about 1000 V/(pC·mm·m) decreasing to less than 1 % at the second bunch and with a long-time level below 0.1 %.

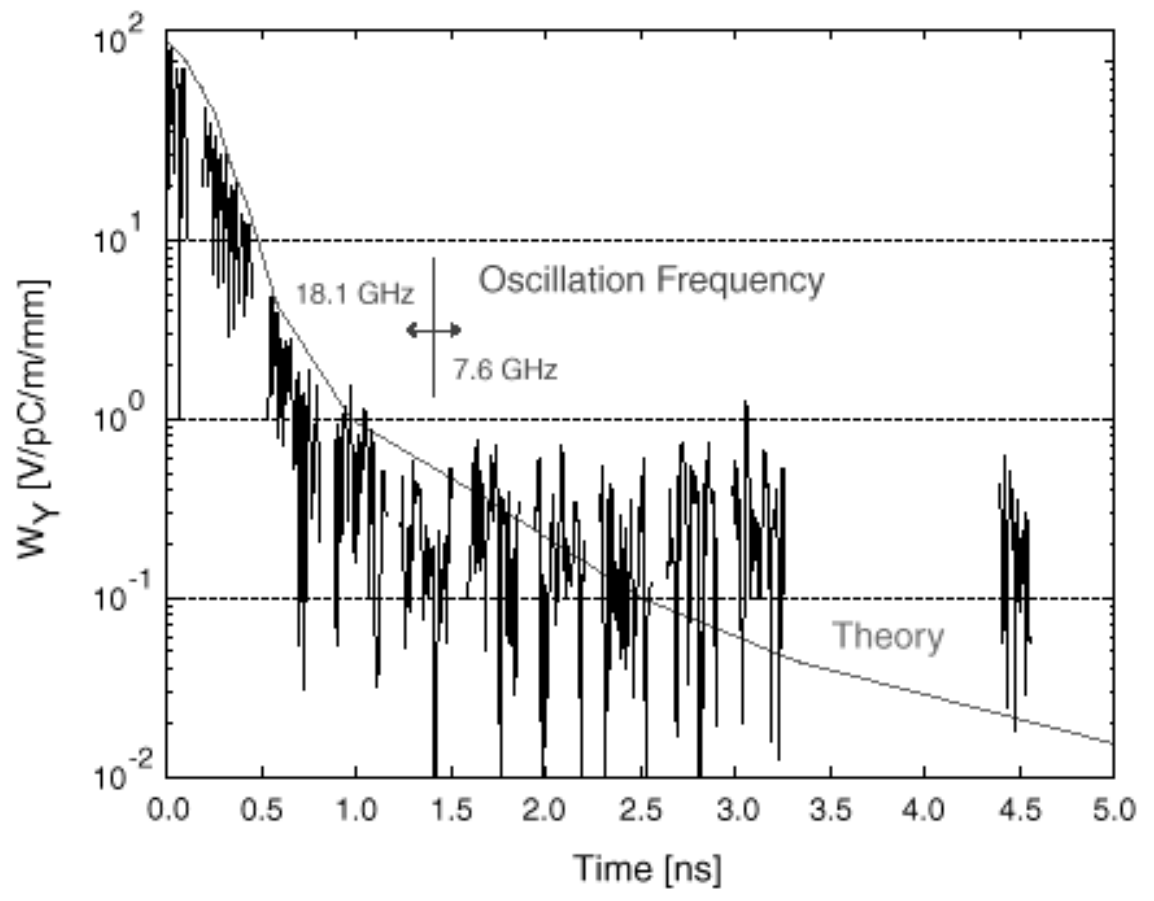

**Figure 6 : Comparison of measured ASSET wakefield levels and theory.**

A 15 GHz scale model of this structure has been tested in the ASSET Test Facility at SLAC. The measured wakefield levels shown in Fig. 6 are in excellent agreement with the theoretical predictions [9]. The 7.6 GHz signal at the 1% level is not related to the structure but to a beam-pipe/structure transition. The recent observation of surface damage at relatively low accelerating gradients (~65 MV/m) and with short pulses (16 ns) in these high group velocity structures is a cause of concern. It is not clear at the moment whether this can be attributed to the geometry of the structure or to other contributing factors such as vacuum level or conditioning procedure.

# 4 THE RF POWER SOURCE

The overall layout of the CLIC RF power source scheme for a 3 TeV centre-of-mass collider is shown in Fig. 7. The RF power for each 624 m section of the main linac is provided by a secondary low-energy high-intensity electron beam which runs parallel to the main linac. The power is generated by passing this electron beam through energy-extracting RF structures in the so-called “Drive Beam Decelerator”.

For the 3 TeV c.m. collider there are 44 drive beams (22 per linac). Each drive beam has an energy of 1.18 GeV and consists of 1952 bunches with a spacing of 2 cm and a maximum charge per bunch of 16 nC. These 22 drive beams, spaced at intervals of 1248 m, are produced as one long pulse by one of the two drive beam generators. By initially sending this drive beam train in the opposite direction to the main beam, different time slices of the pulse can be used to power separate sections of the main linac.

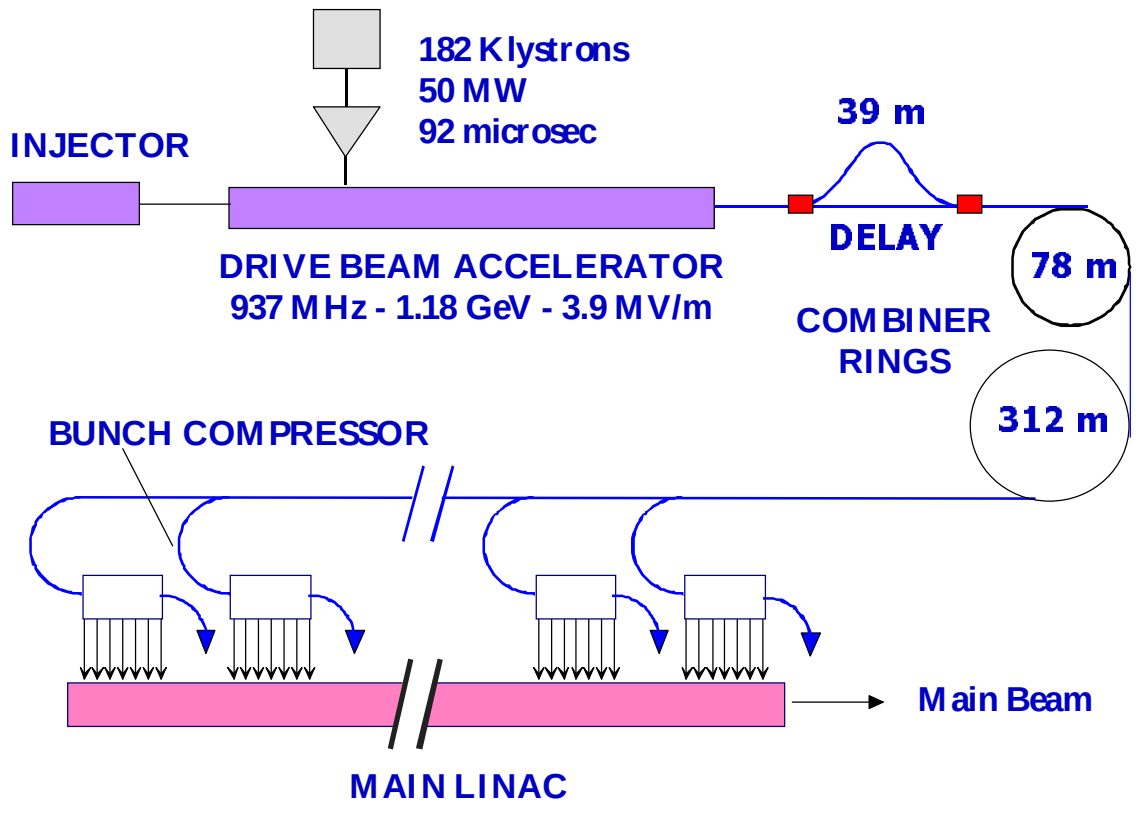

**Figure 7 : Layout of RF power source**

The drive beam is generated as follows [10]. All the bunches (for 22 drive beams) are first generated and accelerated with a spacing of 64 cm as one long continuous train in a normal-conducting fully-loaded 937 MHz linac operating at a gradient of 3.9 MV/m. This 7.5 A 92 μs continuous beam can be accelerated with an RF/beam efficiency ≈ 97%. After acceleration the continuous train of 42944 bunches is split up into 352 trains of 122 bunches using the combined action of a delay line and a grouping of bunches in odd and even RF buckets. These trains are then combined in a 78 m circumference ring by interleaving four successive bunch trains over four turns to obtain a distance between bunches at this stage of 8 cm. A second combination using the same method is subsequently made in a similar, larger 312 m circumference ring, yielding a final distance between bunches of 2 cm. The power-extracting structures consist of four periodically-loaded rectangular waveguides coupled to a circular beam pipe. Each 80 cm long structure provides 458 MW of 30 GHz RF power, enough to feed two accelerating structures. For stability in the drive beam decelerator, these structures have to be damped to reduce long-range transverse wakefield effects.

Two drive-beam accelerator options are presently being studied: to use re-circulation to reduce the installed RF power, and to use a single accelerator to produce the drive beams for both the e- and e+ linacs.

## 5 MAIN BEAM INJECTORS

The main beam injector complex is located centrally (see Fig.1). To reduce cost the same linacs accelerate both electrons and positrons on consecutive RF pulses. The positrons are produced by standard technology already in use at the SLC (SLAC Linear Collider) but with improved performance due to the larger acceptance of the L-band capture linac [11]. The electron and positron beams are damped transversely in specially designed damping rings for low emittances [12]. The damping rings are made up of arcs based on a Theoretical Minimum Emittance (TME) lattice and straight sections equipped with wigglers. The positrons are pre-damped in a pre-damping ring. A specific design for a 3 TeV collider is underway but has not yet been completed. The aim is to provide normalised emittances of $5\times10^{-7}$ and $5\times10^{-9}$ rad.m. in the horizontal and vertical planes respectively, at the entrance to the main linac. The bunches are compressed in two stages in magnetic chicanes [13], the first one after the damping ring using 3 GHz structures, the second one just before injection into the main linac with 30 GHz structures.

An option to use a common injector linac for both main beams and drive beams is being studied.

## 6 THE BEAM DELIVERY AND INTERACTION POINT

Studies of the beam delivery section consisting of a final-focus chromatic correction section and a collimation section for the 3 TeV collider have only just started and for the moment there is no consistent design. A large crossing angle (20 mrad total) is required [14] to suppress the multi-bunch kink instability created by parasitic collisions away from the main interaction point (IP). This however means that crab-cavities will have to be used to avoid a reduction in luminosity. Although the final-focus design is at a very preliminary stage, an optics has been found (see Fig.8) which looks promising [15]. It consists of horizontal and vertical chromatic correction sections followed by a final transformer. 80% of the ideal luminosity is obtained for a 1% full-width flat energy spread of the beams. The rms spot sizes in both planes are 20-30% larger than expected from the simple calculation using the emittance and the beta function at the IP. Peak beta functions reach 1000 km. The length per side is 3.1 km. The design allocation for the total beam delivery section (final focus plus collimation) is $2\times5$ km.

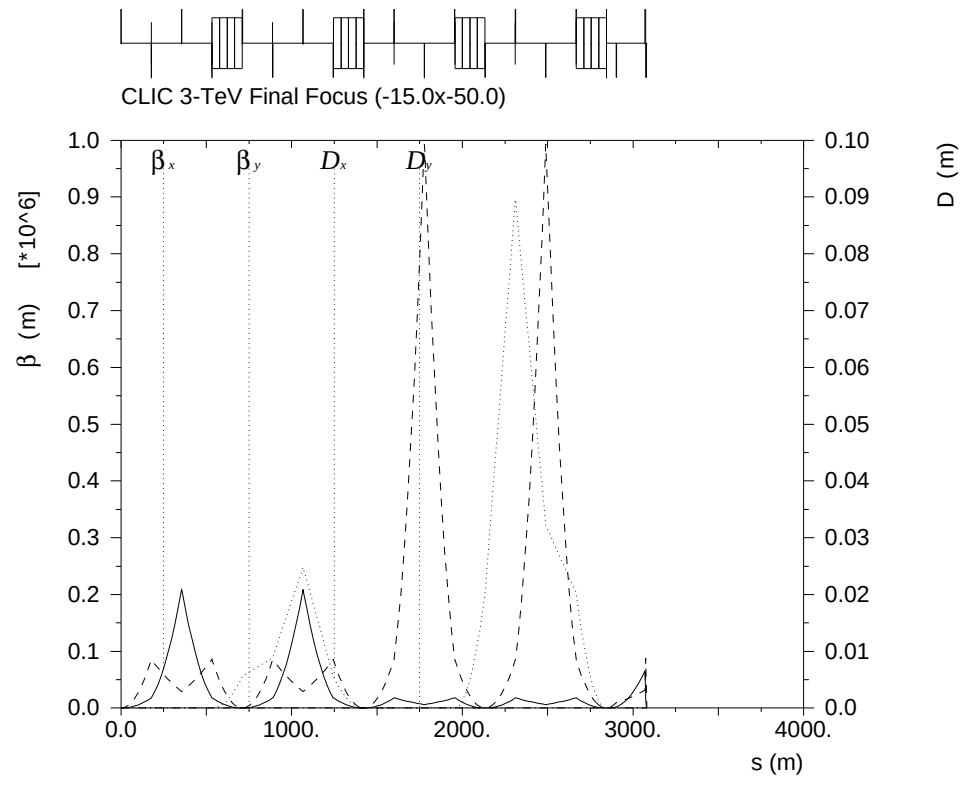

**Figure 8: Final focus optics for a 3 TeV collider.**

The feasibility of maintaining 1 nm beam sizes in collision in the presence of ground movement and component jitter has to be investigated. The use of position feedback systems within the 130 ns pulse are being considered [16].

The consequences of the large mean energy loss (31%) and energy spread (100%) which are produced by the very strong beam-beam forces at the IP in the vertical plane (even when the beams miss each other by 10-20$\sigma$) must be carefully studied. Extraction of a spent beam with 100% energy spread and with a large beam divergence is a concern and will make bending and focussing without beam loss particularly challenging.

The total energy carried by the $6.8\times10^{8}$ coherent pairs at 3 TeV is about 40 Joules, and extracting the particles without producing losses in the detector will be a challenge. Background levels in the detector may also dramatically increase due to their sheer number.

## 7 TEST FACILITIES

The first CLIC Test Facility (CTF1) operated from 1990 to 1995 and demonstrated the feasibility of two-beam power generation. 76 MW of 30 GHz peak power was extracted from a low-energy high-intensity beam and this power was used to generate a gradient in the 30 GHz structure of 94 MV/m for 12 ns.

A second test facility (CTF2) [17] is now being operated. The 30 GHz part of this facility is equipped with a few-microns-precision active-alignment system. The 48-bunch 450 nC drive beam train is generated by a laser-driven S-band RF gun with a $Cs_2Te$ photo-cathode. The beam is accelerated to 40 MeV by two travelling-wave sections operating at slightly different frequencies to provide beam loading compensation along the train. After bunch compression in a magnetic chicane, the bunch train passes through four power extraction and transfer structures, each of which powers one 30 GHz accelerating section (except the third which powers two) with 16 ns long pulses. The single probe beam bunch is generated by an RF gun with a CsI+Ge photo-cathode. It is pre-accelerated to 50 MeV at S-band before being injected into the 30 GHz accelerating linac. The drive beam RF gun has produced a single bunch of 112 nC and a maximum charge of 755 nC in 48 bunches. The maximum charge transmitted through the 30 GHz modules was 450 nC. A series of cross-checks between drive beam charge, generated RF power, and main beam energy gain have shown excellent agreement. The maximum RF power generated by one 0.5 m structure was 27 MW. The highest average accelerating gradient was 59 MV/m and the energy of an 0.7 nC probe beam has been increased by 55 MeV. Unexpected surface damage was found at these field levels and further studies are needed to find the cause. The tests were made under particularly bad vacuum conditions which for the moment makes vacuum a prime suspect. Extremely high gradients were obtained [18] by powering a 30 GHz single-cell resonant cavity directly by the drive beam. The cavity operated without breakdown at a peak accelerating gradient of 290 MV/m. When pushed further, the cavity started to breakdown at surface-field levels around 500 MV/m. The breakdown manifested itself as a field extinction of the decaying pulse at different times in the pulse. At the end of the test when the cavity was breaking down continuously, surface field levels as high as 750 MV/m were obtained.

A new facility (CTF3 – see Fig.9) is under study [19] in collaboration with LAL (France), LNF (Italy) and SLAC (USA), which would test all major parts of the CLIC RF power scheme. To reduce cost, it is based on the use of 3 GHz klystrons and modulators recuperated from the LEP Injector Linac (LIL).

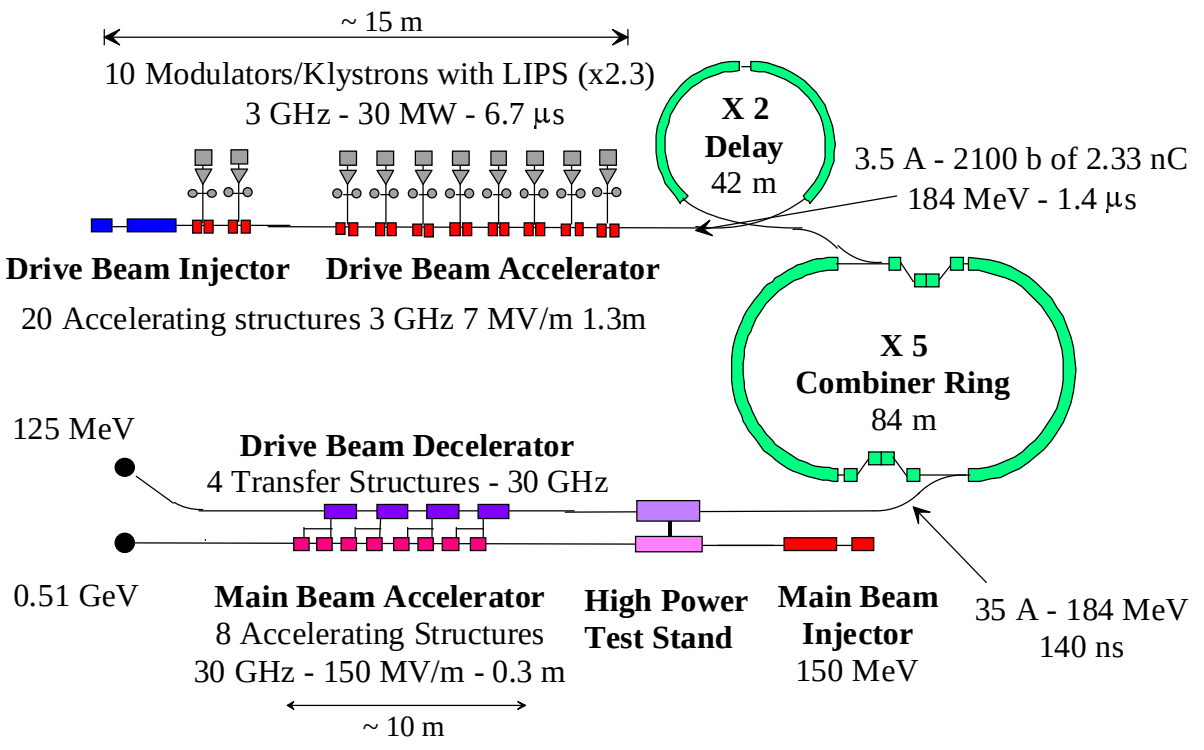

**Figure 9: CTF3 schematic layout.**

The drive beam is generated by a thermionic gun and is accelerated by twenty 1.3 m long fully-loaded 3 GHz structures operating at 7 MV/m with an RF-to-beam efficiency of 96%. The power is supplied by ten 30 MW klystrons and compressed by a factor 2.3 to give a peak power at each structure of 69 MW. The beam pulse is 1.4 µs long with an average current of 3.5 A. The bunches are initially spaced by 20 cm (two 3 GHz buckets) but after two stages of frequency multiplication they have a final spacing of 2 cm. This bunch train, with a maximum charge of 2.3 nC per bunch, is then decelerated by four 0.8 m long transfer structures in the 30 GHz drive beam decelerator from 184 MeV to 125 MeV. Each transfer structure provides 458 MW. The main beam is accelerated from 150 MeV to 510 MeV by eight 30 GHz accelerating structures operating at a gradient of 150 MV/m.

## 8 CONCLUSION

The CLIC Two-Beam scheme is an ideal candidate for extending the energy reach of a future high-luminosity linear collider from 0.5 TeV up to 5 TeV c.m. The high operating frequency (30 GHz) should allow the use of high accelerating gradients (150 MV/m) which shorten the linacs (27.5 km for 3 TeV) and reduce the cost. This level of gradient however has yet to be demonstrated, and the recent unexpected structure damage at much lower field levels is a cause for concern. The effects of the high transverse wakefields have been compensated by a judicious choice of bunch length, charge and focusing strength, such that the emittance blow-up is made independent of the frequency of the accelerating system for equivalent beam trajectory correction techniques. The two-beam RF power source based on a fully-loaded normal-conducting low-frequency linac and frequency multiplication in combiner rings is an efficient, cost-effective and flexible way of producing 30 GHz power. The feasibility of two-beam power production has been demonstrated in the CLIC Test Facilities (CTF1 and CTF2). A third test facility is being studied to demonstrate the newly-proposed drive beam generation and frequency multiplication schemes.